\documentclass[10pt,twocolumn,letterpaper]{article}

\usepackage[pagenumbers]{cvpr} 
\usepackage{multirow} 
\usepackage{hyperref}       
\usepackage{url}            
\usepackage{booktabs}       
\usepackage{amsfonts}       
\usepackage{nicefrac}       
\usepackage{microtype}      
\usepackage{xcolor}         

\usepackage{graphicx}
\usepackage{booktabs}

\usepackage{multirow}
\usepackage{pifont}
\usepackage{threeparttable}
\usepackage{amsmath}
\usepackage{enumitem}

\usepackage{algorithm}
\usepackage{algpseudocode}

\PassOptionsToPackage{dvipsnames}{xcolor}

\usepackage{graphicx}
\usepackage{booktabs}

\usepackage{multirow}
\usepackage{pifont}
\usepackage{threeparttable}
\usepackage{amsmath}
\usepackage{enumitem}

\usepackage{algorithm}
\usepackage{algpseudocode}

\definecolor{cvprblue}{rgb}{0.21,0.49,0.74}

\usepackage[accsupp]{axessibility}

\def\paperID{9667} 
\def\confName{CVPR}
\def\confYear{2025}

\title{Towards Practical Real-Time Neural Video Compression}

\author{Zhaoyang Jia$^1\thanks{ This work was done when Zhaoyang Jia and Linfeng Qi were full-time interns at Microsoft Research Asia. }\quad\!$ Bin Li$^2\quad\!$ Jiahao Li$^2\quad\!$ Wenxuan Xie$^2\quad\!$ Linfeng Qi$^{1*} \quad\!$ Houqiang Li$^1 \quad\!$ Yan Lu$^2$\\
$^1$ University of Science and Technology of China $\ \ ^2$ Microsoft Research Asia\\
{\tt\small \{jzy\_ustc, qlf324\}@mail.ustc.edu.cn, lihq@ustc.edu.cn}\\
{\tt\small \{libin, li.jiahao, wenxie, yanlu\}@microsoft.com}
}

\begin{document}

\newlist{myitemize}{itemize}{1}
\setlist[myitemize,1]{label=\textbullet,leftmargin=5.5mm}

\maketitle

\renewcommand{\thefootnote}{\fnsymbol{footnote}}
\setcounter{footnote}{2}
\footnotetext{This paper is the outcome of an open-source project started from Dec. 2023.}
\renewcommand{\thefootnote}{\arabic{footnote}}
\setcounter{footnote}{1}

\begin{abstract}
We introduce a practical real-time neural video codec (NVC) designed to deliver high compression ratio, low latency and broad versatility. In practice, the coding speed of NVCs depends on 1) computational costs, and 2) non-computational operational costs, such as memory I/O and the number of function calls. While most efficient NVCs prioritize reducing computational cost, we identify operational cost as the primary bottleneck to achieving higher coding speed. Leveraging this insight, we introduce a set of efficiency-driven design improvements focused on minimizing operational costs. Specifically, we employ implicit temporal modeling to eliminate complex explicit motion modules, and use single low-resolution latent representations rather than progressive downsampling. These innovations significantly accelerate NVC without sacrificing compression quality. Additionally, we implement model integerization for consistent cross-device coding and a module-bank-based rate control scheme to improve practical adaptability. Experiments show our proposed DCVC-RT achieves an impressive average encoding/decoding speed at 125.2/112.8 fps (frames per second) for 1080p video, while saving an average of 21\% in bitrate compared to H.266/VTM. The code is available at \url{https://github.com/microsoft/DCVC}.
\end{abstract}    
\section{Introduction}
\label{sec:introduction}

\begin{figure}[t]
  \centering
    \includegraphics[width=\linewidth]{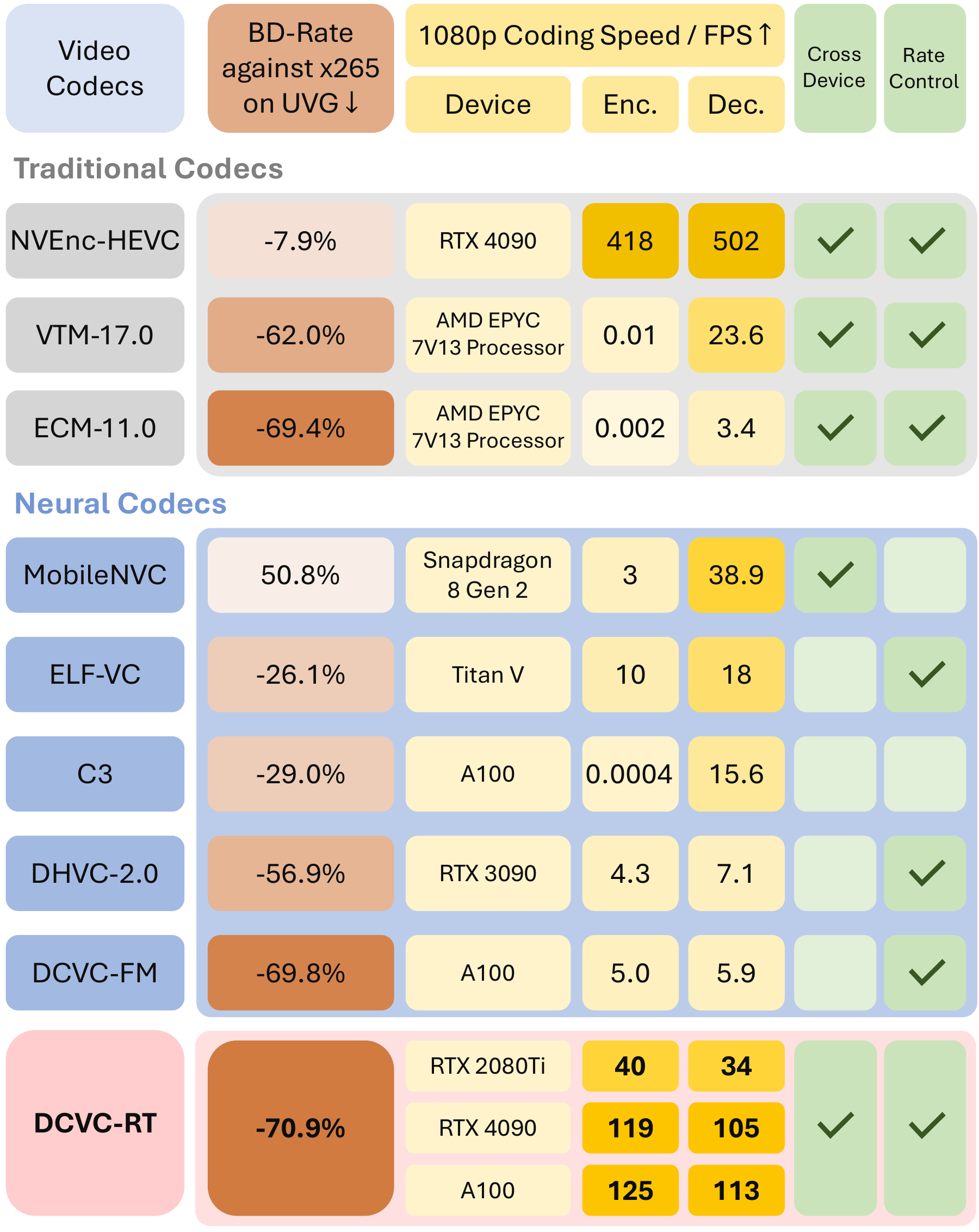}
    \vspace{-2mm}
    \caption{Towards practical real-time neural video codecs (NVCs). 
    Recent advanced NVCs have demonstrated either excellent rate-distortion performance, or improved versatility like integrated cross-device coding consistency or rate-control capabilities.
    In this paper, we further address the core obstacles of achieving real-time coding to close the last mile toward a practical NVC solution.
    Our DCVC-RT not only achieves state-of-the-art compression ratio but is also deployable on consumer devices for real-time video coding.} 
  \label{fig:practical_compare}
  \vspace{-2mm}
\end{figure}

\begin{figure*}[t]
  \centering
    \includegraphics[width=\linewidth]{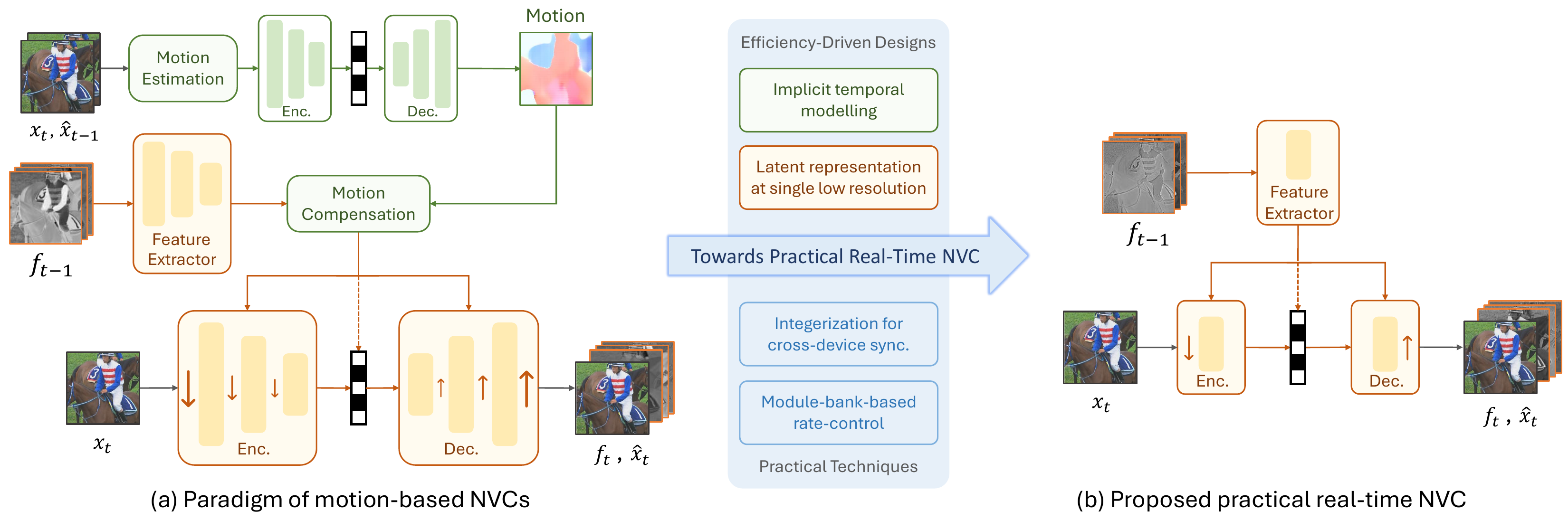}
    \vspace{-3mm}
    \caption{Paradigm shift. To enhance efficiency, we eliminate explicit motion-related modules and adopt implicit temporal modeling. We also propose learning latent representations at a single low resolution, replacing the traditional progressive downsampling approach. Additionally, DCVC-RT supports integerization for cross-device consistency and incorporates a module-bank-based rate-control mechanism.
    } 
  \label{fig:Paradigm}
  \vspace{0mm}
\end{figure*}

Neural video codecs (NVCs) have exhibited significant potential in reducing redundancy within video data to achieve higher compression ratios. Since the early work DVC \cite{DVC}, substantial advances \cite{lu2020end, yang2020learning, hu2022coarse, agustsson2020scale, DCVC-FM, shi2022alphavc, liu2023mmvc, CANF-VC, kim2023c3, lu2024high, qi2024long, van2021overfitting, gao2024pnvc, hu2023complexity} have been made in enhancing the rate-distortion performance of NVCs. Recent NVCs have surpassed traditional codecs like H.265/HM \cite{HM}, H.266/VTM \cite{VTM}, and ECM \cite{ECM}. In this context, compression ratio is no longer the primary bottleneck for NVCs. Instead, the key challenge now lies in how to make NVCs more practical and deployable for real-world applications, to effectively utilize the advantages of such compression ratios.

In response, recent efforts have concentrated on enhancing the functionality and versatility of NVCs. Tian et al. \cite{tian2023towards} introduced auxiliary calibration bitstream transmission to improve cross-device coding accuracy, while MobileCodec \cite{le2022mobilecodec} and MobileNVC \cite{van2024mobilenvc} employ deterministic integer calculations to ensure consistent output across different devices. For rate-control functionality, methods like ELF-VC \cite{rippel2021elf}, DCVC-FM \cite{DCVC-FM}, and DHVC \cite{lu2024deep, lu2024high} offer controllable rate adjustment within a single model. Zhang et al. \cite{zhang2023neural} developed a rate allocation network for precise bitrate control. These innovations have significantly improved the practicality of NVCs, bringing them closer to real-world deployment.

Despite these advancements, a critical challenge persists for practical NVCs: how to effectively accelerate NVCs for real-time coding? Actually, current NVCs struggle to balance coding speed with rate-distortion performance, leading to a suboptimal \textbf{rate-distortion-complexity} trade-off. For instance, while MobileNVC \cite{van2024mobilenvc} achieves real-time decoding on consumer hardware, its compression ratio is even lower than x264 \cite{ffmpeg}. C3 \cite{kim2023c3} provides efficient decoding but relies on time-consuming optimization-based encoding. DHVC-2.0 \cite{lu2024high} requires pipelining across multiple GPUs (e.g., 4) to achieve real-time decoding, but its efficiency drops significantly in typical single-GPU environments. Although these methods have significantly accelerated NVCs, real-time coding of 1080p video with high compression ratios on consumer devices remains elusive.

In this paper, we aim to address the core obstacles of achieving real-time coding to close the last mile toward a practical NVC solution. To accelerate NVCs, our first step is to rethink the complexity problem. While most existing research focuses on reducing \textbf{computational complexity}, typically measured by the number of multiply-accumulate operations (MACs) during model inference, this alone does not determine the actual coding speed. In practice, many other operations like communication between hardware components, also significantly impact performance. For instance, auto-regressive entropy models \cite{minnen2018joint, DCVC, kim2023c3} require frequent function calls, which incur significant time overhead despite low overall computational cost. Additionally, memory I/O costs of tensors increase with larger tensor sizes, even at the same computational load. We define these factors as the \textbf{operational complexity}. Surprisingly, our findings show that high operational overhead, rather than computational cost, is the primary bottleneck in accelerating NVCs. 

Based on this insight, we propose a new perspective to accelerate NVCs by reducing operational complexity. In this process, we preserve model capacity by prioritizing more computational capability on the most critical modules while eliminating less essential ones. Firstly, we remove complex motion estimation and compensation process, significantly cutting down the number of components to directly lower the operation frequency. The saved computational capacity is reallocated to frame coding modules to achieve more effective implicit temporal modelling. Additionally, we propose learning latent representations at a single low resolution, i.e., 1/8 of the original image size. Compared to commonly used progressive downsampling method, this approach greatly reduces latent-wise memory I/O overhead while facilitating more effective latent transformations, leading to improved rate-distortion-complexity performance. 

With the aforementioned real-time innovations, we further implement model integerization to ensure cross-device consistency and introduce a module-bank-based rate-control technique. Together, these advancements culminate in a practical real-time NVC, DCVC-RT. As shown in Fig. \ref{fig:practical_compare}, it enables 1080p coding on consumer GPUs like the NVIDIA RTX 2080Ti with an average speed of 40 fps for encoding and 34 fps for decoding. On an NVIDIA A100 GPU, it reaches an impressive 125 fps for encoding and 113 fps for decoding. Compared to VTM/H.266, our model provides a 21.0\% bitrate reduction when using the challenging single intra-frame setting. Additionally, it matches the compression ratio of the advanced DCVC-FM \cite{DCVC-FM} while delivering over 18 times faster coding speed. To the best of our knowledge, DCVC-RT is the first practical NVC to achieve real-time coding with a high compression ratio on consumer hardware.

We summarize the contributions of this paper as follows:
\begin{itemize}
\item We investigate the complexity challenges in NVCs and identify operational complexity, rather than computational complexity, as the primary bottleneck.
\item Based on this insight, we propose several efficiency-driven designs to reduce operational complexity and enable real-time NVCs. We further enhance the functionality to introduce a practical real-time NVC, DCVC-RT.
\item To the best of our knowledge, DCVC-RT is the first real-time NVC to achieve high rate-distortion performance, enabling 1080p real-time coding on consumer hardware with a $21\%$ bitrate reduction compared to VTM/H.266.
\end{itemize}

\section{Related Works}
\label{sec:related}

Since the introduction of DVC \cite{DVC}, most research on neural video codecs (NVCs) has focused on enhancing rate-distortion performance. By advancing temporal modeling capabilities \cite{lin2020m, agustsson2020scale, shi2022alphavc, pourreza2023boosting, alexandre2023hierarchical, DCVC-TCM, qi2023motion}, improving latent distribution estimation \cite{DCVC-HEM, DCVC-DC, liu2023mmvc, CANF-VC}, and refining coding paradigms \cite{liu2020conditional, ladune2021conditional, DCVC, chen2023neural, mentzer2022vct, lu2024deep}, the compression efficiency of NVCs has seen substantial improvement. Recent state-of-the-art NVCs \cite{DCVC-FM, qi2024long} now outperform top traditional codecs like ECM \cite{ECM}. For NVCs, the primary challenge has shifted from rate-distortion optimization to enhancing functionality and adaptability for real-world deployment.

\textbf{Real-Time Coding.} While real-time coding has been explored in neural image codec \cite{minnen2018joint, wang2023EVC, liu2023learned, yang2023computationally, zhang2024gaussianimage, jia2024generative}, it remains relatively underexplored in neural video codecs. Some efforts \cite{rippel2021elf, tian2023towards} aim to reduce computational complexity for faster coding but still fall short of achieving real-time 1080p performance. INR-based methods \cite{kim2023c3, hu2023complexity, gao2024pnvc} focuses on efficient decoding but requires a time-consuming optimization process for encoding. DHVC-2.0 \cite{lu2024high} uses multi-GPU pipelines to achieve real-time throughput for decoding, but falls short in meeting real-time latency requirements and on more common single-GPU devices. MobileNVC \cite{van2024mobilenvc} achieves real-time decoding throughput but only matches the compression ratio of x264 \cite{ffmpeg}. In this paper, we present a novel perspective to reduce operational cost rather than computational cost in NVC. Based on it, we introduce several key efficiency-driven techniques to simultaneously achieve 1080p real-time latency with a compression ratio comparable to ECM.

\textbf{Practical Functionality.} For video codecs, maintaining calculation consistency across different devices is a crucial functionality. Typically, this inconsistency arises from nondeterministic floating-point calculations. Ball{\'e} et al. \cite{balle2018integer} and He et al. \cite{he2022post} introduced model integration into neural image codecs to enforce deterministic integer calculations. Similarly, MobileCodec \cite{le2022mobilecodec} and MobileNVC \cite{van2024mobilenvc} implement integration in their NVCs to ensure cross-device consistency. Recently, Tian et al. \cite{tian2023towards} proposed eliminating inconsistency by introducing auxiliary calibration bitstreams. Another important aspect of video codecs is their rate-control capability, particularly in scenarios such as streaming or real-time communication. Zhang et al. \cite{zhang2023neural} developed a rate allocation network to precisely manage bitrate. Other approaches \cite{rippel2021elf, DCVC-FM, lu2024deep} enable continuous, controllable bitrates within a single model, adjusting the model to the target bitrate by manipulating the quantization parameters (qp). However, these methods fail to achieve both cross-device consistency and rate-control capability simultaneously. In contrast, we introduce a practical NVC that support both capabilities, along with real-time coding and a high compression ratio.
\section{Rethink the Complexity Problem in NVCs}
\label{sec:complexity}

\begin{figure*}[t]
  \centering
    \includegraphics[width=\linewidth]{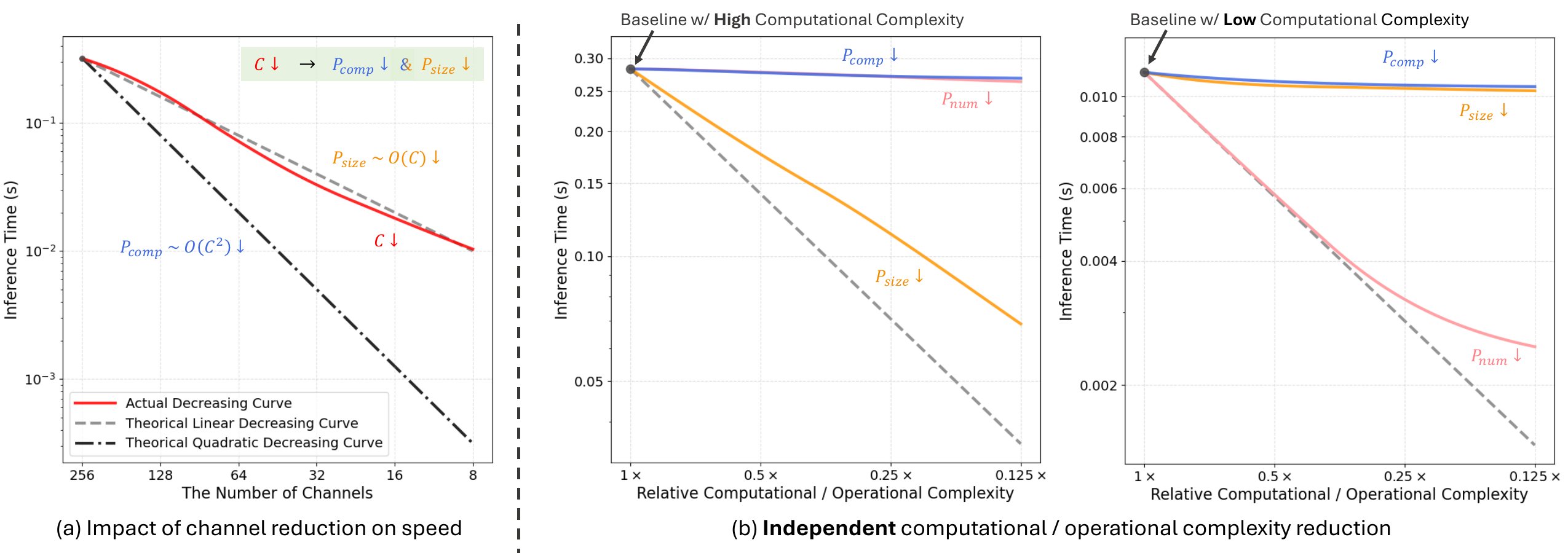}
    \vspace{-3mm}
    \caption{Analysis on \textcolor[HTML]{2F58DD}{computational complexity $P_{comp}$} and operational complexity, including \textcolor[HTML]{E48302}{latent representation size $P_{size}$} and \textcolor[HTML]{FF6970}{number of modules $P_{num}$}. 
    (a) Reducing channels results in a quadratic decrease in \textcolor[HTML]{2F58DD}{$P_{comp}$}, yet inference time decreases almost linearly, indicating that computational cost is not the primary speed bottleneck. (b) We independently reduce one of \textcolor[HTML]{2F58DD}{$P_{comp}$}, \textcolor[HTML]{E48302}{$P_{size}$} and \textcolor[HTML]{FF6970}{$P_{num}$} to identify the main factors affecting time cost. Results show that \textcolor[HTML]{E48302}{$P_{size}$} is most critical at high computational complexity, while the \textcolor[HTML]{FF6970}{$P_{num}$} is more significant at low computational complexity.
    } 
  \label{fig:Complexity_Analysis}
  \vspace{-3mm}
\end{figure*}

The primary challenge in the practical application of existing NVCs is their low coding speed. While recent efforts have aimed at reducing computational costs \cite{kim2023c3, tian2023towards, lu2024deep}, achieving real-time acceleration remains elusive. To address this, we conducted experiments to rethink the complexity problem in NVC acceleration.

In CNNs, \textbf{computational complexity} $P_{comp}$ are typically dominated by matrix multiplications, often mitigated by reducing the channel $C$, as it scales computational complexity by $O(C^2)$. However, our findings reveal that reducing $C$ does not result in the expected quadratic speedup. As illustrated in Fig. \ref{fig:Complexity_Analysis} (a), speed improves in a more linear fashion as $C$ decreases. This suggests that factors other than computational complexity are limiting the coding speed.

In practice, numerous factors influence coding speed. We identify two key factors: 1) \textbf{latent representation size} $P_{size}$, which primarily influences memory I/O costs of latent tensors; and 2) \textbf{number of modules} $P_{num}$, which affects the total operation counts and the overhead of function calls. These factors primarily influence additional operations beyond hardware computations, which we term \textbf{operational complexity}, distinct from computational complexity. We conducted experiments that independently controlled $P_{comp}$,  $P_{size}$ and $P_{num}$ to observe their impact on inference speed. The independent control of each factor is achieved by balancing the number of modules $N$, channel $C$ and latent resolution $H\times W$.  For example, halving $C$ while doubling $H \times W$ maintains a constant $P_{size}$ but reduces $P_{comp}$ due to its quadratic relationship with $C$. 

The results in Fig. \ref{fig:Complexity_Analysis} (b) reveal several key insights. 1) Operational complexity, rather than computational complexity, is the main speed bottleneck. Reducing computational costs without addressing operational factors leads to only marginal improvements in inference time. This also explains why reducing channels results in a linear, rather than quadratic, decrease in time—since the latent size decreases linearly with the number of channels. 2)  When computational complexity is high, latent representation size becomes the dominant limiting factor. When computational complexity is low, the number of modules becomes the key bottleneck. This suggests that different parts of the model require different optimization strategies.

These insights offer a new perspective on accelerating NVCs by reducing operational complexity. Typically, lowering computational complexity leads to diminished compression performance. However, since computational complexity is no longer the primary speed bottleneck, we can focus on lowering operational complexity while preserving computational capacity. In our design, we prioritize computational capability to the most critical modules while eliminating less essential ones, which ensures sufficient model capacity and achieves a better rate-distortion-complexity trade-off.

\begin{figure*}[t]
  \centering
    \includegraphics[width=\linewidth]{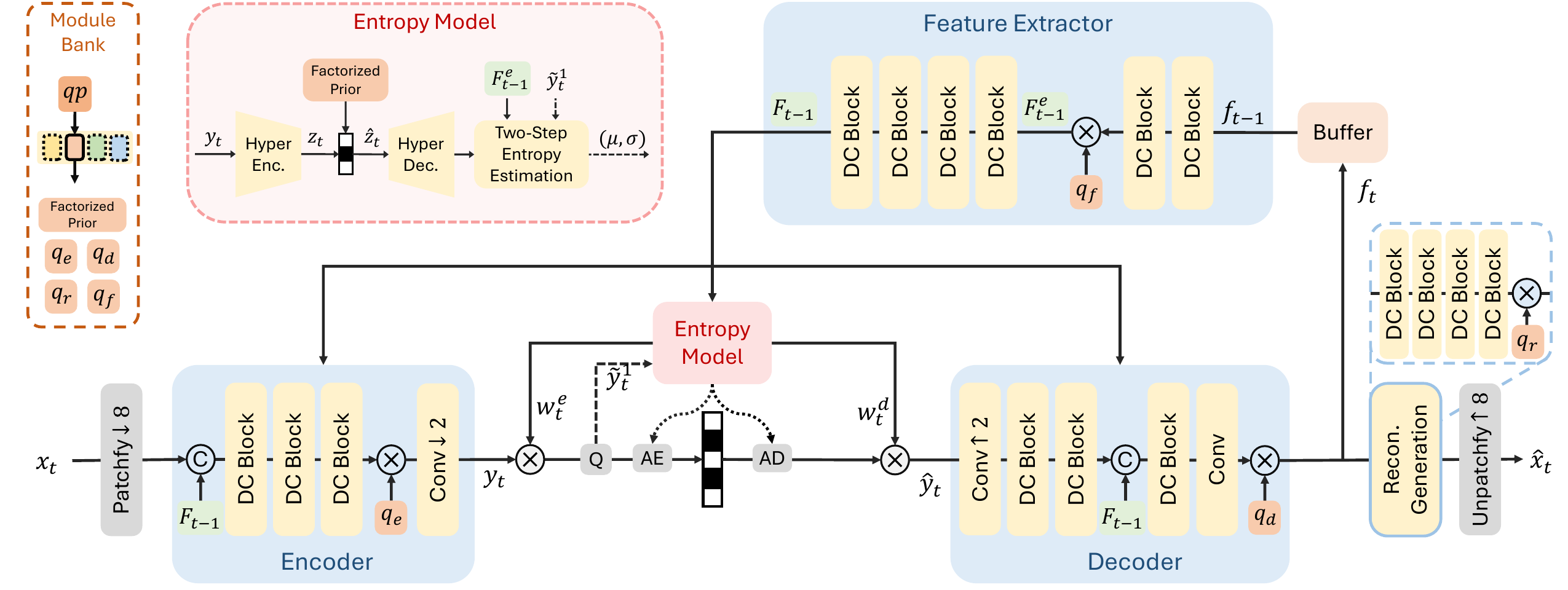}
    \vspace{-1mm}
    \caption{Framework overview. DC Block, Q, AE and AD represent depth-wise convolution block, quantization, arithmetic encoder and decoder, respectively. $F_{t-1}$ and $F^e_{t-1}$ are temporal contexts extracted from previously decoded latent $f_{t-1}$. Frames are transformed into latents at 1/8 resolution using patch embedding \cite{dosovitskiy2021an}, and key modules such as the encoder, decoder, frame extractor, and reconstruction generation operate at this single scale for efficient feature learning. DCVC-RT eliminates explicit motion modeling, resulting in a streamlined design with drastically reduced operational complexity and real-time performance.
    } 
    \vspace{-1mm}
  \label{fig:Framwork}
\end{figure*}
\section{Towards Practical Real-Time NVC}
\label{sec:methods}

\subsection{Overview}
\label{sec:methods-overview}

\vspace{-0.5mm}

The framework of the proposed DCVC-RT is illustrated in Fig. \ref{fig:Framwork}. To compress current frame $x_t$, we first patchfy it into $\frac{1}{8}$-scale latents using patch embedding \cite{dosovitskiy2021an}. Then we perform conditional coding \cite{DCVC, CANF-VC, DCVC-FM} in this single low resolution (Section \ref{sec:methods-single}) to achieve efficient coding. During extracting the temporal context information, DCVC-RT incorporates implicit temporal modeling (Section \ref{sec:methods-implicit}) to avoid complex motion-estimation-motion-compensation process. To improve versatility, we introduce a module-bank-based rate-control method (Section \ref{sec:methods-rate}) and enable model integerization (Section \ref{sec:methods-int}) for cross-device consistency. 

\subsection{Latents at Single Low Resolution}
\label{sec:methods-single}

Originating from the concept of compressive auto-encoders \cite{theis2017lossy}, most NVCs progressively downsample latents to reduce the dimension. At each layer, they downsample the latent by half while doubling the number of channels. It enables a comparable computational capacity $P_{comp}$ across layers,
\begin{equation}
    P_{comp} \sim O((2C)^2\cdot H / 2\cdot W / 2)=O(C^2\cdot H\cdot W)
\end{equation}
while the latent size $P_{size}$ is gradually reduced
\begin{equation}
    P_{size} = 2C\cdot H / 2\cdot W / 2=1 / 2\cdot C\cdot H\cdot W
\end{equation}

\begin{figure*}[t]
    \centering
    \includegraphics[width=\linewidth]{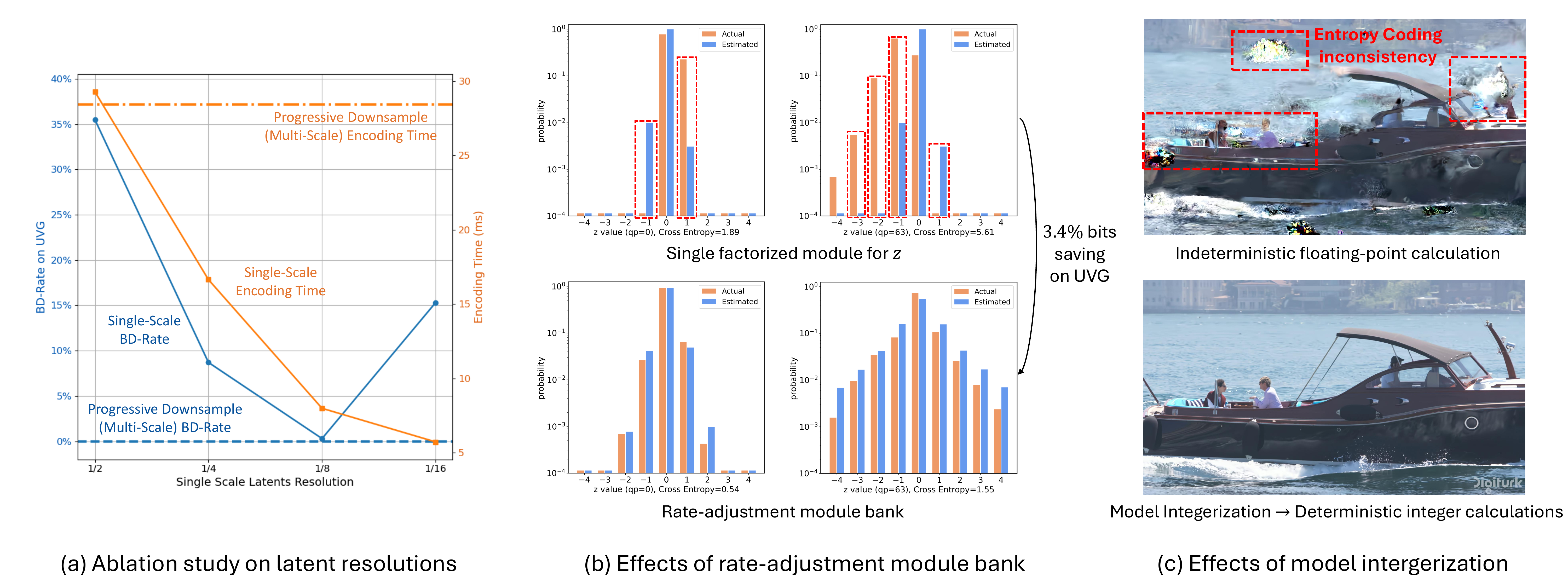}
  \vspace{-4mm}
    \caption{Analysis of different components. (a) Ablation study on learning latent representations at a single resolution. All models maintain equal computational complexity (MACs) for fairness. (b) Example of probability estimation of $z$. Using a module bank instead of a single factorized module achieves an average bit savings of 3.4\%. (c) Cross platform coding test. We perform encoding on an NVIDIA A100 GPU, while decoding uses an RTX 2080Ti. Model integerization effectively eliminates coding inconsistencies across platforms.}
    \label{fig:methods-main}
  \vspace{-2mm}
\end{figure*}

In Section. \ref{sec:complexity}, we learn that the latent size can be the main bottleneck for coding speed. From this operational complexity perspective, a question arises: can we learn latents at a single low resolution to eliminate the high operational costs associated with a large $P_{size}$? To explore this, we directly downsample frames to a single scale using patch embedding and apply conditional coding to compress them at the same scale. Results in Fig. \ref{fig:methods-main} (a) prove the feasibility of this method, where learning single low scale latents notably boosts encoding speed. For example, learning latents at 1/8 scale is about $3.6\times$ faster than progressive downsampling.

While reducing latent scales accelerate model inference, it also impacts rate-distortion performance. Although computational capacity is maintained across scales, the varied receptive field may influence the performance. At a single high 1/2 resolution, the restricted receptive field results in notable performance degradation. However, as scales decrease, the receptive field expands significantly, and at 1/8 scale, it even surpasses the receptive field of progressive downsampling. This extended receptive field is essential for enhancing temporal modeling and reducing temporal redundancy, leading to a comparable BD-Rate of $0.3\%$ under the same model capacity. However, a performance drop is observed at $1/16$ scale. In our model, $1/16$ scale latents with $C=512$ yields a latent size of $512\cdot H/16\cdot W/16=2\cdot H\cdot W$, which is even smaller than the original frame $3\cdot H\cdot W$. It significantly limits representative capacity and degrades the compression ratio. In contrast, at the $1/8$ scale, using $C=256$ yields a sufficient latent size of $4 \cdot H \cdot W$. Considering these factors, we adopt 1/8 single-scale latent learning.

\subsection{Implicit Temporal Modelling}
\label{sec:methods-implicit}

In video coding, temporal correlation modeling is crucial for effective redundancy reduction. Most existing NVCs achieve this by an explicit motion estimation and motion compensation process. Typically, motion coding needs low computational complexity since motions are simpler and easier for compression. However, we observe that existing motion modules usually use a high number of module layers. For example, we observe that the motion coding branch in \cite{DCVC-FM} exhibits $13\times$ lower computational complexity than the conditional coding branch (74 kMACs per pixel versus 932 kMACs per pixel), despite having up to half as many convolutional layers (123 layers versus  225 layers). As discussed in Section \ref{sec:complexity}, this high number of layers increases operational complexity, becoming the primary speed bottleneck for low-computational-complexity motion modules.

To address this, DCVC-RT adopts implicit temporal modeling, extracting temporal context using a single and simple feature extractor instead of complex motion-based temporal context extraction. Technically, this temporal context is concatenated with the current latent along the channel dimension, allowing the encoder-decoder to process them jointly for redundancy reduction. By eliminating the need for motion estimation and compensation, the number of modules is directly reduced to lower the operation frequency, significantly enhancing the coding speed. 

\textbf{Analysis on Different Motion Contents.} We compare implicit and explicit modeling across different motion types for a more comprehensive evaluation.As shown in Tab. \ref{tab:aba-motion}, implicit modeling slightly improves BD-Rate by 0.4\% on small motions while showing a modest 3.2\% reduction on large motions. Nonetheless, with a 3.4$\times$ faster encoding time, implicit modeling is a more practical solution for real-time applications. Additionally, it surpasses explicit motions in scene changes scenarios, since scene change cannot be effectively modeled by motions. These results highlight its advantages in the rate-distortion-complexity trade-off.

\begin{table*}[t]
    \caption{Ablation study on implicit temporal modelling. We compare it with using explicit motions under different motion contents. The motion range is measured using a pretrained SEA-RAFT \cite{wang2025sea}. Further details are provided in the supplementary material. }
    \vspace{-1mm}
    \centering
    \setlength{\tabcolsep}{6pt}
    \resizebox{0.86\linewidth}{!}{
	\begin{tabular}{ l | c  c c  c | c}
        \toprule
		\multirow{2}{*}{Temporal Modelling} & \multicolumn{4}{c|}{BD-Rate} & \multirow{2}{*}{Encoding Time}\\
		~ & MCL-JCV Average & Large Motion & Small Motion & Scene Change  & ~\\
	    \midrule
        Explicit Motions   & \textbf{0.0\%} & \textbf{0.0\%} & 0.0\% & 0.0\% & 27.2 ms (3.4$\times$)\\
        Implicit Temporal Modelling & 2.1\% & 3.2\% & \textbf{--0.4\%} & \textbf{--4.7\%} & 8.0 ms (1$\times$)\\
        \bottomrule
	\end{tabular}
	}
  \label{tab:aba-motion}
  \vspace{-1mm}
\end{table*}

\subsection{Module-Bank-Based Rate Control}
\label{sec:methods-rate}

In DCVC-RT, rate control is achieved through variable-rate coding with dynamic rate adjustment. While existing variable-rate codecs \cite{rippel2021elf, DCVC-FM} primarily focus on adjusting the distribution of latent $y$, they typically compress hyper information $z$ using a single factorized prior module. Since $z$ generally accounts for less than $1\%$ of the total bits, it has minimal impact on their performance. However, in DCVC-RT, we find that $z$ contributes over 10\% bits of $y$ on average, since the absence of motion bits makes $z$ critical in spatial-temporal modeling. In this case, inaccurate distribution estimation for $z$ severely affects overall performance.

To address this, we introduce a rate-adjustment module bank (shown in the top left of Fig. \ref{fig:Framwork}). It learns a range of hyperprior modules model varied distributions across different quantization parameter (qp). As shown in Fig. \ref{fig:methods-main} (b), this module bank closely aligns estimated distributions with actual distributions, achieving about 3\% bit savings. Extending this approach, we further introduce separate vector banks for different modules (e.g., $q_e$, $q_d$, $q_f$ and $q_r$ for encoder, decoder, feature extractor, and reconstruction network, respectively). Each vector bank is designed to learn a set of vectors that adaptively scale the latent representations based on their characteristics, enabling flexible and fine-grained amplitude adjustments. DCVC-RT achieves efficient rate control using this module bank, with the rate-control results provided in the supplementary material.

Moreover, DCVC-RT supports hierarchical quality control by adjusting qp offsets for different frames. Compared to \cite{DCVC-FM} that employs separate feature adaptors for this purpose, our method achieves improved consistency with rate adjustment and enhanced flexibility in practical applications.

\begin{table*}[t]
    \caption{BD-Rate (\%) comparison in YUV420 colorspace. All frames with intra-period=--1.}
    \vspace{-2mm}
    \centering
    \renewcommand\arraystretch{1.4}
    \setlength{\tabcolsep}{3.5pt}
    \resizebox{0.86\linewidth}{!}{
    \begin{threeparttable}
    	\begin{tabular}{ l c c c c c c c c c }
            \toprule
    		& \multirow{2}{*}{UVG} & \multirow{2}{*}{MCL-JCV} & \multirow{2}{*}{HEVC B} & \multirow{2}{*}{HEVC C} & \multirow{2}{*}{HEVC D} & \multirow{2}{*}{HEVC E} & \multirow{2}{*}{Average} & \multicolumn{2}{c}{Coding Speed} \\
            & ~ & ~ & ~ & ~ & ~ & ~ & ~ & Enc. & Dec. \\
            \hline
            VTM-17.0    & $0.0$   & $0.0$   & $0.0$   & $0.0$   & $0.0$   & $0.0$   & $0.0$  & $0.01$ fps & 23.6 fps\\
            \hline
            HM-16.25    & $40.1$  & $48.6$  & $47.6$  & $41.0$  & $34.5$  & $42.8$  & $42.4$  & $0.05$ fps & 39.6 fps\\
            \hline
            ECM-11.0    & $-20.0$ & $-22.1$ & $-22.2$  & $-21.2$  & $-20.4$  & $-17.2$  & $-20.5$  & $0.002$ fps & 3.4 fps  \\
            \hline
            DCVC-DC     & $6.5$   & $-4.4$  & $13.1$  & $-3.4$  & $-14.8$ & $90.2$ & $14.5$ & $3.3$ fps & $4.3$ fps \\
            \hline
            DCVC-FM     & $-17.6$ & $-8.4$ & $-15.7$ & $-30.2$ & $-37.6$ & $-23.0$ & $-22.1$ & $3.4$ fps & $4.2$ fps  \\
            \hline
            DCVC-FM (fp16)     & $-16.8$ & $-8.0$ & $-15.4$ & $-30.2$ & $-37.5$ & $-20.2$ & $-21.3$ & $5.0$ fps & $5.9$ fps  \\      
            \hline
            \textbf{DCVC-RT} (fp16)     & $-24.0$  & $-14.8$   & $-16.6$ & $-21.0$  & $-27.3$ & $-22.4$ & $-21.0$ & $125.2$ fps & $112.8$ fps \\
            \bottomrule
    	\end{tabular}
        \begin{tablenotes}
            \item Note: Some values differ slightly from those in \cite{DCVC-FM} since we use actual BPP instead of estimated BPP. 
        \end{tablenotes}
    \end{threeparttable}
	}
  \label{tab:compare_yuv_allf}
  \vspace{-1mm}
\end{table*}

\begin{figure*}[t]
  \centering
    \includegraphics[width=\linewidth]{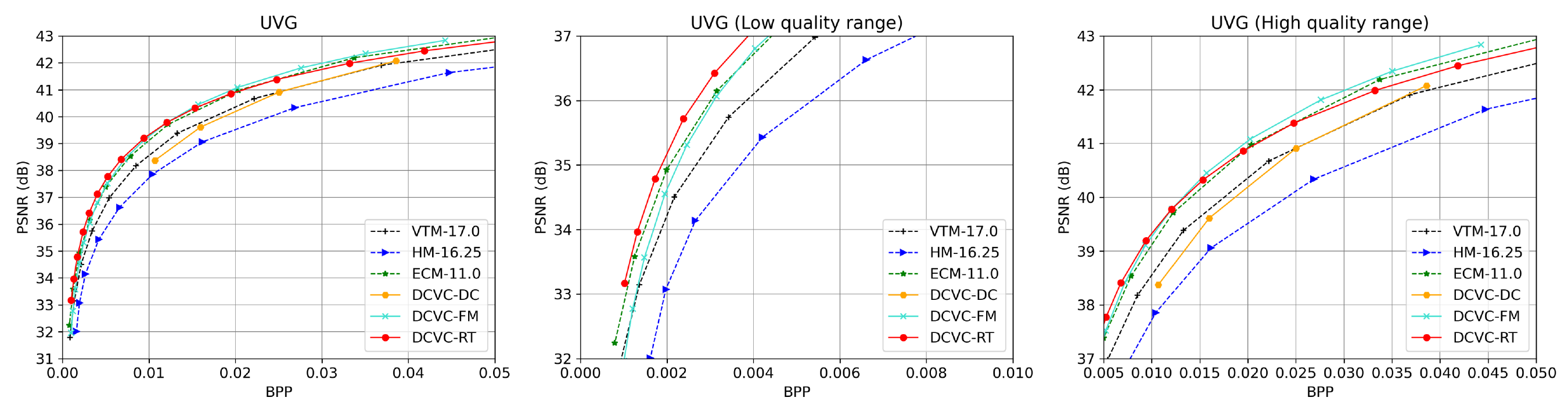}
    \vspace{-6mm}
    \caption{Rate-distortion curves for UVG. All frames are tested in YUV420 colorspace with intra-period=--1. Results on more datasets are in the supplementary materials. } 
  \label{fig:RD-Curve}
  \vspace{-3mm}
\end{figure*}

\subsection{Model integerization}
\label{sec:methods-int}

For NVCs, the indeterminism of floating-point calculations can cause inconsistencies when distributing video content. To address this, we implement 16-bit model integerization. This approach enables deterministic integer calculations and ensures consistent output across different devices. More concretely, the equation between a floating-point feature value $v_f$ and an int16 value $v_i$ is as follows
\begin{equation}
v_i = \text{round}(K_1\cdot v_f)
\label{equ:float-to-int}
\end{equation}
We set $K_1=512$, such that the floating-point value $1.0$ is mapped to $512$ in int16, and given that the valid int16 range is $[-32768, 32767]$, the corresponding range of floating-point values is $[-64.0, 63.998]$. We observe this is sufficiently large to represent the values during model inference. We set the accumulator data type to int32 in convolutional kernels, and no overflow issues have been observed. Besides convolutions and basic arithmetic operations, we adopt a precomputed lookup table to handle the nonlinear Sigmoid function, that maps an arbitrary int16 value to its corresponding output. Through this training-free model integerization, DCVC-RT can perform deterministic integer calculations, ensuring cross-device consistency. An example is shown in Fig. \ref{fig:methods-main} (c), with further results in the supplementary material.

\section{Experiments}
\label{sec:experiments}

\subsection{Settings}
\label{sec:experiments-settings}

\textbf{Datasets}. We use Vimeo-90k \cite{vimeo} to train DCVC-RT with 7-frame sequences, and process the original Vimeo videos \cite{ori_vimeo} to create longer sequences for fine-tuning by following  \cite{DCVC-FM}. We evaluate DCVC-RT on HEVC Class B$\sim$E \cite{flynn16common}, UVG \cite{uvg}, and MCL-JCV \cite{mcl-jcv}.\\
\textbf{Test Details}. For traditional codecs, we compare with HM \cite{HM}, VTM \cite{VTM} and ECM\cite{ECM}, which represent the best H.265, H.266 encoder and the prototype of next generation traditional codec, respectively. Detailed configurations are provided in the supplementary material. For neural codecs, we compare with advanced NVCs including DCVC-DC \cite{DCVC-DC} and DCVC-FM \cite{DCVC-FM}. Following \cite{DCVC-FM}, we test all frames with an intra-period of --1 on YUV420 and RGB colorspace. We conduct all tests under low delay conditions. Rate-distortion performance is assessed by the BD-Rate \cite{bd-rate}. Additionally, we note that many existing NVCs compare with traditional codecs using the estimated entropy, which is unfair as they overlook header information. In this paper, we ensure a fair comparison by retesting them with actual binary bit-streams that include necessary header information. By default, coding speed is tested on a single NVIDIA A100 GPU with an AMD EPYC 7V13 processor. We measure the average latency across different quantization parameters (qp) on a resolution of $1920 \times 1080$.
\vspace{1mm}
\\
\textbf{Training Details}. To accommodate variable rates within a single model, we randomly assign different qp between $[0, 63]$ in each training iteration. In a group of 8 pictures, the qp offset is set to $[0, 8, 0, 4, 0, 4, 0, 4]$ for hierarchical quality. We follow \cite{DCVC-DC} to adopt a hierarchical weight setting for the distortion term to support a hierarchical quality structure. The corresponding  $\lambda$ values are interpolated between $1$ and $768$, following the same method as in \cite{DCVC-FM}. We use the combined distortion loss in both YUV and RGB colorspace \cite{DCVC-FM} to support both colorspace in a single model. 

\begin{table*}[t]
    \caption{Complexity analysis. The encoding / decoding speed (measured in frames per second, fps) are evaluated across various resolutions and devices, including the NVIDIA A100, NVIDIA A6000, RTX 4090, and RTX 2080 Ti. Average BD-Rate results are presented using VTM as the anchor. MACs are tested on 1080p videos. OOM indicates out-of-memory conditions.}
    \vspace{-1mm}
    \centering
    \setlength{\tabcolsep}{6pt}

    \begin{tabular}{cc}
        \begin{minipage}{0.47\textwidth} 
            \centering
            \resizebox{0.84\linewidth}{!}{
        	\begin{tabular}{ l | c | c | c}
                \toprule
        		Model & Average BD-Rate & MACs & Params\\
        	    \midrule
                DCVC-DC & 14.5\% & 2642G & 19.8M \\
        	    \midrule
                DCVC-FM (fp16) & --21.3\% & 2642G & 18.3M \\
        	    \midrule
                DCVC-RT (fp16) & --21.0\% & 385G & 20.7M \\
        	    \midrule
                DCVC-RT (int16) & --18.3\% & 385G & 20.7M \\
                \bottomrule
        	\end{tabular}
            }
            \subcaption{Computational complexity and BD-Rate.} 
            \vspace{3px}
        \end{minipage}
        & 
        \hspace{-15px}
        \begin{minipage}{0.52\textwidth}
            \centering
            \resizebox{\linewidth}{!}{
        	\begin{tabular}{ l | c | c | c | c}
                \toprule
        		Model & A100 & A6000 & 4090 & 2080Ti\\
        	    \midrule
                DCVC-DC & 3.3 / 4.3 & 1.7 / 2.2 & 2.3 / 2.9 & 0.8 / 1.4 \\
        	    \midrule
                DCVC-FM (fp16) & 5.0 / 5.9 & 3.1 / 3.8 & 3.7 / 4.4 & 1.9 / 2.3 \\
        	    \midrule
                DCVC-RT (fp16) & 125.2 / 112.8 & 70.4 / 63.8 & 118.8 / 105.3 & 39.5 / 34.1  \\
        	    \midrule
                DCVC-RT (int16) & 28.3 / 20.9 & 23.4 / 17.5 & 52.3 / 38.8 & 18.4 / 13.4 \\
                \bottomrule
        	\end{tabular}
            }
            \subcaption{Coding speed on $1920\times1080$ videos.} 
            \vspace{3px}
        \end{minipage} \\
        
        \hspace{-15px}
        \begin{minipage}{0.47\textwidth}
            \centering
            \resizebox{\linewidth}{!}{
        	\begin{tabular}{ l | c | c | c | c}
                \toprule
        		Model & A100 & A6000 & 4090 & 2080Ti\\
        	    \midrule
                DCVC-DC & 0.8 / 1.0 & 0.4 / 0.5 & OOM & OOM \\
        	    \midrule
                DCVC-FM (fp16) & 1.0 / 1.2 & 0.6 / 0.7 & OOM & OOM \\
        	    \midrule
                DCVC-RT (fp16) & 35.5 / 29.5 & 18.5 / 16.2 & 29.9 / 26.5 & 11.6 / 9.9 \\
        	    \midrule
                DCVC-RT (int16) & 7.3 / 5.2 & 6.1 / 4.4 & 12.5 / 9.5 & 4.4 / 3.2 \\
                \bottomrule
        	\end{tabular}
            }
            \subcaption{Coding speed on $3840\times2160$ videos.} 
        \end{minipage}
        & 
        \hspace{-15px}
        \begin{minipage}{0.52\textwidth}
            \centering
            \resizebox{\linewidth}{!}{
        	\begin{tabular}{ l | c | c | c | c}
                \toprule
        		Model & A100 & A6000 & 4090 & 2080Ti\\
        	    \midrule
                DCVC-DC & 6.5 / 7.9 & 3.5 / 4.3 & 5.5 / 6.7 & 2.1 / 2.9 \\
        	    \midrule
                DCVC-FM (fp16) & 8.5 / 9.4 & 5.9 / 6.6 & 9.3 / 10.4 & 4.0 / 4.7 \\
        	    \midrule
                DCVC-RT (fp16) & 173.9 / 149.2 & 147.3 / 132.5 & 225.1 / 185.2 & 73.3 / 67.0 \\
        	    \midrule
                DCVC-RT (int16) & 51.7 / 39.2 & 49.5 / 38.1 & 105.2 / 81.1 & 37.0 / 25.8 \\
                \bottomrule
        	\end{tabular}
            }
            \subcaption{Coding speed on $1280\times720$ videos.} 
        \end{minipage}
    \end{tabular}
  \label{tab:compare_complexity}
  \vspace{-3mm}
\end{table*}

\subsection{Comparison Results}
\vspace{-1mm}
\label{sec:experiments-settings}
In Tab. \ref{tab:compare_yuv_allf}, we present the BD-rate comparison for the YUV420 format under all frame intra period --1 settings. As depicted in the table, DCVC-RT achieves an average 21.0\% bits saving compared to VTM, which is slightly better than 20.5\% of ECM. It showcases comparable compression ratio to the advanced NVC DCVC-FM with an impressive $25$ times faster encoding speed, reaching 125.2 fps for encoding and 112.8 fps for decoding. This demonstrate the superior performance of DCVC-RT in term of rate-distortion-complexity trade-off. In the RGB colorspace, DCVC-RT achieves a 14.0\% bits saving compared to VTM, closely matching the 15.8\% savings of DCVC-FM. Detailed results are provided in the supplementary material.

Fig. \ref{fig:RD-Curve} presents the rate-distortion curve on UVG. DCVC-RT showcases better performance with VTM across the entire quality range. Particularly in the low-quality range ($<0.02$ bpp), DCVC-RT exhibits the best performance. However, there is a performance drop in the high-quality range. This drop can be attributed to the lightweight model design adopted in DCVC-RT, resulting in reduced model capability compared to larger models. Notably, this drop mainly occurs above $40$ dB, where human vision struggles to distinguish between different qualities. In the supplementary material, we further examine the compression performance of DCVC-RT as model capacity increases. Our large model achieves the highest compression ratio across all bitrate ranges while maintaining real-time performance.

\subsection{Complexity Analysis}
\label{sec:experiments-speed}

Tab. \ref{tab:compare_complexity} presents the complexity analysis. Compared to DCVC-DC and DCVC-FM, DCVC-RT achieves significantly lower computational complexity while maintaining a comparable compression ratio. Coding speed is evaluated across multiple input resolutions and GPU devices, consistently demonstrating at least a 20$\times$ speed improvement. On the A100 GPU, DCVC-RT (fp16) reaches real-time 4K 30fps coding, while on consumer-grade devices like the RTX 2080 Ti, it achieves 1080p 30fps coding. These results highlight the efficiency of DCVC-RT across diverse conditions.

\subsection{Integerization Results}
\label{sec:experiments-int}

Our model supports 16-bit integer calculations. As shown in Tab. \ref{tab:compare_complexity}, our integerization strategy introduces minimal impact on compression performance, with DCVC-RT (int16) still outperforming VTM by 18.3\%. Dataset-specific BD-Rate results are in the supplementary material. In terms of coding speed, DCVC-RT (int16) achieves 1080p 30 fps coding on an RTX 4090 and 720p 24 fps coding on an RTX 2080Ti.

However, we observe a significant slowdown in coding speed when using int16 mode compared to fp16. This is mainly because most modern GPUs lack dedicated optimization for int16 operations. This difference is particularly pronounced on the A100 GPU, where highly optimized Tensor Cores make fp16 processing over four times faster than int16. Although int16 mode theoretically has the potential to enable faster inference than fp16, we anticipate that future hardware developments and engineering will help bridge this performance gap.

\section{Conclusion and Limitation}
\label{sec:conclusion}
In this paper, we propose a practical, real-time neural video codec (NVC) focused on high compression ratio, low latency, and broad versatility. By analyzing the complexity of NVCs, we identify operational cost, rather than computational cost, as the primary bottleneck to coding speed. Based on this insight, we employ implicit temporal modeling and a single low-resolution latent representation, which significantly accelerates processing without compromising compression quality. Additionally, we introduce model integerization for consistent cross-device coding and a module-bank-based rate control scheme to enhance practical adaptability. As far as we known, DCVC-RT is the first NVC achieving 110 fps coding on 1080p video with a 21\% bitrate savings compared to H.266/VTM. DCVC-RT serves as a notable landmark in the journey of NVC evolution.

While DCVC-RT supports int16 mode, its coding speed remains slower than fp16 due to limited hardware optimization for int16 inference. In the future, we hope this can be solved by further hardware optimization and engineering.

{
    \small
    \bibliographystyle{ieeenat_fullname}
    \bibliography{main}
}

\clearpage
\appendix
\noindent{\huge \textbf{Appendices}}\\

\noindent This document provides the supplementary material to our proposed practical real-time neural video codec (NVC), i.e. DCVC-RT.
\section{Test Settings}
\label{sec:sup-test}

For a fairer comparison with traditional codecs, we employ their best settings to represent their best compression ratio. We test them in both YUV420 and RGB colorespaces for a comprehensive comparison.

\textbf{YUV420 colorspace}. We focus our comparison on the YUV420 colorspace, which is commonly optimized in traditional video codecs. Our evaluation includes the comparison with HM \cite{HM}, VTM \cite{VTM}, and ECM \cite{ECM}, representing the best H.265 encoder, the best H.266 encoder, and the  prototype of the next-generation traditional codec, respectively. For each traditional codec, we utilize the officially provided config files: \textit{encoder\_lowdelay\_main10.cfg}, \textit{encoder\_lowdelay\_vtm.cfg}, and \textit{encoder\_lowdelay\_ecm.cfg}. The parameters for encoding are as:

\begin{myitemize}
    \item
	-c \{{\em config file name}\}\par
	-\/-InputFile=\{{\em input video name}\}\par
	-\/-InputBitDepth=8\par
	-\/-OutputBitDepth=8 \par
	-\/-OutputBitDepthC=8 \par
	-\/-FrameRate=\{{\em frame rate}\}\par
	-\/-DecodingRefreshType=2\par
	-\/-FramesToBeEncoded=\{{\em frame number}\}\par
	-\/-SourceWidth=\{{\em width}\}\par
	-\/-SourceHeight=\{{\em height}\}\par
	-\/-IntraPeriod=\{{\em intra period}\}\par
	-\/-QP=\{{\em qp}\}\par
	-\/-Level=6.2\par
	-\/-BitstreamFile=\{{\em bitstream file name}\}\par
\end{myitemize}

\textbf{RGB colorspace}. In our experiments, the raw videos are stored at YUV420 format. So we convert them from YUV420 to RGB colorspace for testing. Following JPEG AI \cite{jpeg_ai, jpeg_ai2} and \cite{DCVC-DC, DCVC-FM}, we utilize BT.709 to perform this conversion, which brings higher compression ratio compared to the commonly-used BT.601. We test traditional codecs using 10-bit YUV444 as the internal colorspace and evaluate the final results in RGB. \cite{DCVC-DC, DCVC-FM} prove that, for traditional codecs, this brings better compression ratio than the direct measurement in RGB. For HM, VTM, and ECM, we utilize \textit{encoder\_lowdelay\_rext.cfg}, \textit{encoder\_lowdelay\_vtm.cfg}, and \textit{encoder\_lowdelay\_ecm.cfg} as the config file, respectively. The parameters for coding are as:

\begin{myitemize}
    \item
	-c \{{\em config file name}\}\par
	-\/-InputFile=\{{\em input file name}\}\par
	-\/-InputBitDepth=10\par
	-\/-OutputBitDepth=10 \par
	-\/-OutputBitDepthC=10 \par
	-\/-InputChromaFormat=444\par
	-\/-FrameRate=\{{\em frame rate}\}\par
	-\/-DecodingRefreshType=2\par
	-\/-FramesToBeEncoded=\{{\em frame number}\}\par
	-\/-SourceWidth=\{{\em width}\}\par
	-\/-SourceHeight=\{{\em height}\}\par
	-\/-IntraPeriod=\{{\em intra period}\}\par
	-\/-QP=\{{\em qp}\}\par
	-\/-Level=6.2\par
	-\/-BitstreamFile=\{{\em bitstream file name}\}\par
\end{myitemize}

\section{Implementation Details}
\label{sec:sup-impl}

\subsection{Module Structures}
\label{sec:sup-impl-module}

\begin{figure}[t]
  \centering
    \includegraphics[width=0.9\linewidth]{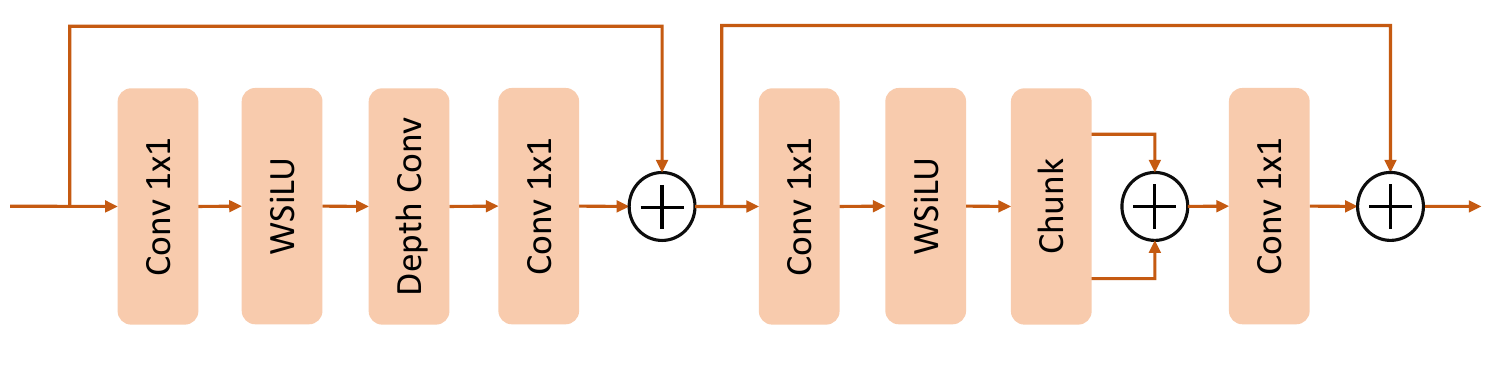}
    \vspace{-2mm}
    \caption{Structures of depth-wise convolution block (DC Block). The formulation of WSiLU is detailed in Section. \ref{sec:sup-impl-module}. Chunk denotes splitting latents into two parts along the channel dimension.}
  \label{fig:Modules}
  \vspace{-2mm}
\end{figure}

DCVC-RT follows a conditional coding manner \cite{DCVC, CANF-VC, DCVC-DC, DCVC-FM}. While the main paper covers the core network structure, we additionally illustrate the architecture of the depth-wise convolution block (DC Block) in Fig. \ref{fig:Modules}, which is not detailed in the manuscript. Here, WSiLU denotes a weighted SiLU function \cite{elfwing2018sigmoid}, formulated as:
\begin{equation}
    \text{WSiLU}(x) = x \cdot \text{Sigmoid}(\alpha \cdot x)
\end{equation}
where the weighting parameter $\alpha$ is set to $4$ by default.

\subsection{Entropy model}
\label{sec:sup-entropy-model}
DCVC-RT adopts a two-step distribution estimation scheme \cite{DCVC-HEM} to balance the coding speed and compression ratio. Although it results in a slight performance drop compared to more complex entropy models \cite{DCVC-DC, minnen2018joint}, we adopt it since it enables significantly faster coding speed.

\begin{figure*}[t]
  \centering
    \includegraphics[width=\linewidth]{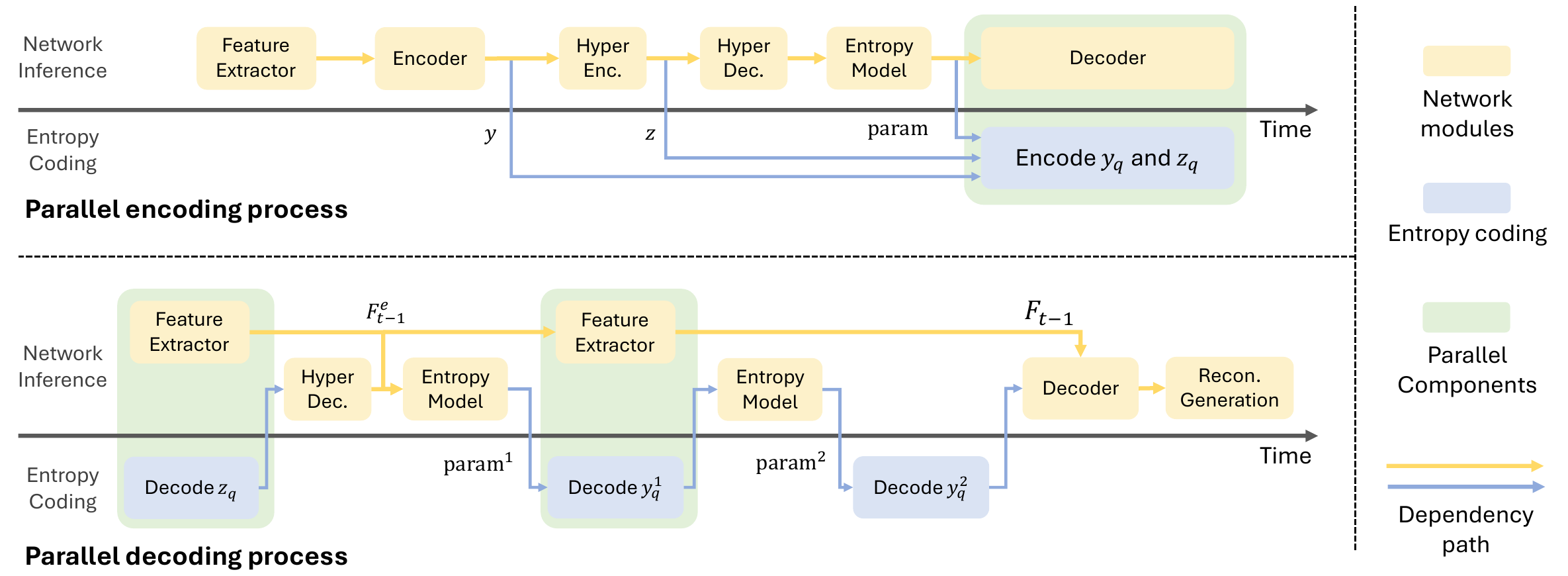}
    \caption{Encoding and decoding pipeline. $y_q$ and $z_q$ represent the symbols to be encoded in entropy coding. This parallel coding approach results in an average \textbf{12\%} speedup in our encoding process and a \textbf{9\%} speedup in our decoding process.} 
  \label{fig:ParallelCoding}
\end{figure*}

\subsection{Parallel Coding}

In practical scenarios, coding speed is influenced by both model inference time and entropy coding time. For DCVC-RT, encoding a 1080p frame typically requires around 7.8 ms for network inference and up to 2.6 ms for entropy coding, indicating that entropy coding significantly contributes to overall latency.

To accelerate the process, we propose a parallel coding scheme, depicted in Fig. \ref{fig:ParallelCoding}. Observing that certain network modules can be inferred independently of entropy coding, we conduct these modules with entropy coding in parallel. This parallelization does not cause severe hardware resource contention, so the latency can be effectively reduced. For instance, we can perform entropy coding on the CPU without affecting concurrent network inference process on the GPU.

In encoding, the feature extractor, encoder, and entropy model are sequentially inferred, generating symbols $y_q$ and $z_q$ for entropy coding and the quantized latents for decoder inference. Since the coding of $y_q$ and $z_q$ does not depend on decoder inference, these processes are parallelized. It is worth noting that, as the reference feature is buffered before reaching the reconstruction module, the reconstruction generation module does not typically need to be inferred. In this case, our encoding is usually faster than decoding.

\begin{table}[t]
    \caption{Sequences used for ablation on implicit temporal modelling. MCL-JCV contains 30 sequences, we denote each sequence using their ID, e.g., 01 denotes sequence videoSRC01\_1920x1080\_30.}
    \vspace{-1mm}
    \centering
    \setlength{\tabcolsep}{6pt}
    \resizebox{\linewidth}{!}{
	\begin{tabular}{ l | l}
        \toprule
		Motion Type & Sequences\\
	    \midrule
		Large Motion & 02, 04, 05, 07, 08, 10, 11, 14, 17, 19, 20, 21, 22, 24, 26\\
	    \midrule
		Small Motion & 01, 03, 06, 09, 12, 13, 15, 16, 18, 23, 25, 27, 28, 29, 30\\
	    \midrule
		Scene Change & 04, 14, 19, 20, 21, 25, 26, 27, 28, 29\\
        \bottomrule
	\end{tabular}
	}
  \label{tab:sup-motion}
  \vspace{-1mm}
\end{table}

\begin{figure}[t]
  \centering
    \includegraphics[width=\linewidth]{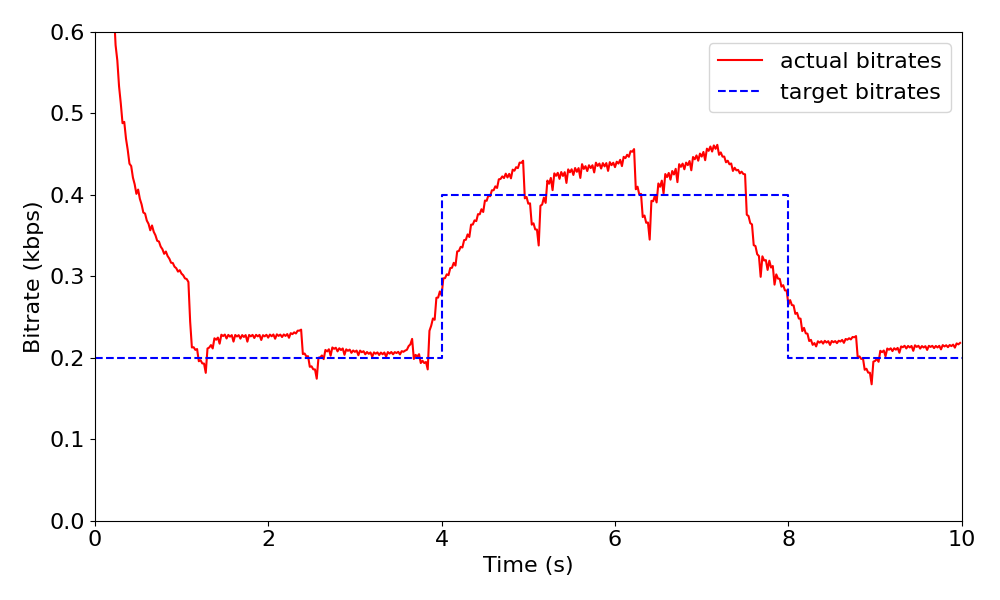}
    \vspace{-2mm}
    \caption{An example for rate-control on \textit{Cactus} sequence. } 
  \label{fig:Rate-Control}
\end{figure}

In decoding, the entropy decoding of $z_q$ is independent of the feature extractor, allowing us to process them concurrently. Since decoding $z_q$ is fast, we only infer the first part of the feature extractor at this stage. After probability estimation, the remainder of the feature extractor is inferred alongside the entropy decoding of $y_q$. Notably, we adopt a two-step coding scheme, where we first decode $y_q^1$ followed by $y_q^2$. Experiments show that the decoding time of $y_q^1$ covers the time needed for feature extractor inference.

In our implementation, the parallel coding approach results in an average 12\% speedup in the encoding process and a 9\% speedup in the decoding process, demonstrating its effectiveness in accelerating DCVC-RT.

\subsection{Model Integerization}
\label{sec:sup-model-int}

We provide a detailed explanation to better understand the model integerization pipeline. As mentioned in Equation \ref{equ:float-to-int}, all int16 features are linearly mapped from its floating-point counterparts in the original model. Moreover, we also create linear mappings for convolution weights $w$ and biases $b$. For a convolution layer, we have
\begin{align*}
    w_i &= \text{round}(K_2\cdot w_f) \\
    b_i &= \text{round}(K_2\cdot b_f)
\end{align*}
where $K_2=8192$ in our implementation. Let $x$ be the input feature and $y$ be the output feature, we derive the integer convolution from the floating-point convolution as follows
\begin{align*}
    y_f &= \text{conv}(x_f, w_f) + b_f \\
    \frac{y_i}{K_1} &= \text{conv}(\frac{x_i}{K_1}, \frac{w_i}{K_2}) + \frac{b_i}{K_2} \\
    y_i &= \frac{\text{conv}(x_i, w_i) + b_i \cdot K_1}{K_2}
\end{align*}
where $K_1=512$ as mentioned in Equation \ref{equ:float-to-int}.

The processes above are rewritten as an algorithmic workflow in Algorithm \ref{alg:conv}.

\begin{algorithm}
    \caption{Model Integerization for a Convolution}
    \label{alg:conv}
    \begin{algorithmic}[1]
        \State \textbf{Input:} Input feature $x$; Convolution weight $w_f$ and bias $b_f$; Hyperparameters $K_1$ and $K_2$.
        \State \textbf{Output:} Output feature $y_i$.
        \If{$x$ is of type fp16} \Comment{The input frame}
            \State $x_i \gets round(K_1\cdot x).\text{to\_int16}()$
        \Else \Comment{The remaining layers}
            \State assert $x$ is of type int16
            \State $x_i \gets x$
        \EndIf
        \State Get int16 weight: $w_i = \text{round}(K_2\cdot w_f).\text{to\_int16}()$
        \State Get int16 bias: $b_i = \text{round}(K_2\cdot b_f).\text{to\_int16}()$
        \State $a \gets \text{conv}(x_i, w_i) + b_i \cdot K_1$
        \State Return: $y_i \gets \text{clip}(\frac{a}{K_2}, -32768, 32767)$
    \end{algorithmic}
\end{algorithm}

\section{Experimental Results}
\label{sec:sup-exp}

\begin{figure}[t]
  \centering
    \includegraphics[width=\linewidth]{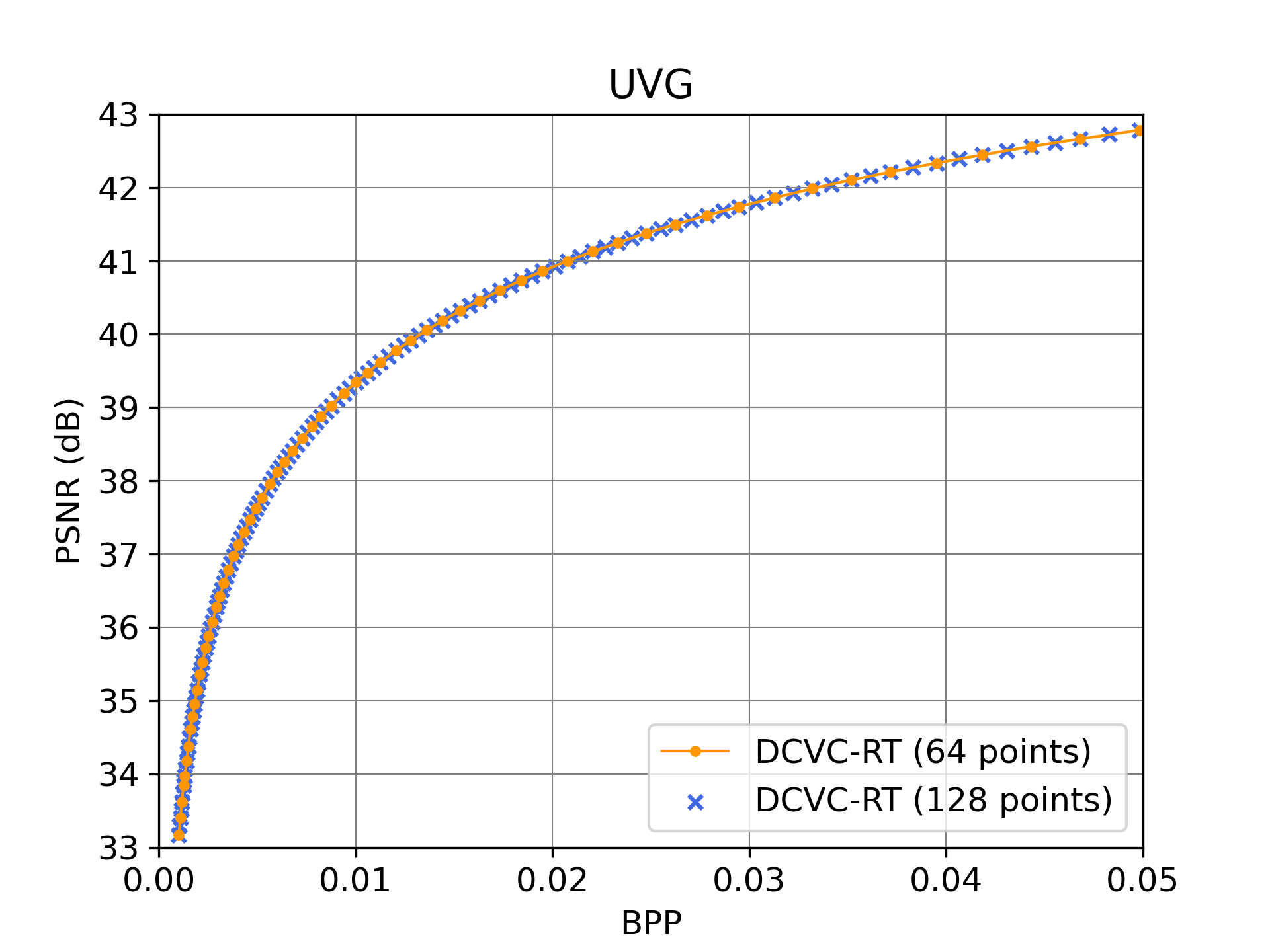}
    \vspace{-2mm}
    \caption{Extend supported rate number from original $64$ rates to $128$ rates using interpolation on UVG dataset. } 
  \label{fig:rate_adjustment_128}
  \vspace{-2mm}
\end{figure}

\subsection{Ablation on Implicit Temporal Modelling}
In Tab. 2. of the main paper, we conduct an ablation study to compare explicit motion estimation with implicit temporal modeling under different motion conditions. Motion amplitude is determined using a pretrained SEA-RAFT \cite{wang2025sea} model to calculate the average motion between consecutive frames, enabling categorization into large or small motion content. Scene changes are identified manually. Tab. \ref{tab:sup-motion} lists the sequences categorized by motion type.

\subsection{Results on Rate-Control}
Rate control is an essential feature for video codecs, particularly for applications like streaming and real-time communication. By adjusting the quantization parameter (qp), DCVC-RT achieves effective rate-control. Fig. \ref{fig:Rate-Control} illustrates this capability, showcasing how DCVC-RT can modulate bitrates effectively.

\subsection{Per-Module Complexity Analysis}
Tab. \ref{tab:per-module-complexity} presents a per-module complexity analysis, detailing the average inference time and the corresponding MACs for each module. Note that since both our encoding and decoding pipeline both do not execute all modules (e.g., reconstruction generation is skipped during encoding), the sum of latencies does not equal the overall latency.

\begin{table}[t]
    \vspace{-2px}
    \centering
    \caption{Breakdown on per-module complexity for coding a 1080p frame on A100 GPU.}
    \vspace{-2px}
    \renewcommand\arraystretch{1.4}
    \setlength{\tabcolsep}{3.5pt}
    \resizebox{0.95\linewidth}{!}{
    \begin{tabular}{l | c | c | c | c | c}
        \midrule
          \multirow{2}{*}{Metric} & \multirow{2}{*}{Encoder} & \multirow{2}{*}{Decoder} & Feature & Entropy & Reconstruction\\
          ~ & ~ & ~ & Extractor & Model & Generation\\
        \midrule
          Latency & 1.4 ms & 1.4 ms & 2.4 ms & 2.4 ms & 2.2 ms\\
        \midrule
          kMACs/pixel & 30.1 & 34.0 & 53.0 & 31.5 & 56.4\\
        \midrule
    \end{tabular}
    \label{tab:per-module-complexity}
	}
    \vspace{-2px}
\end{table}

\begin{figure*}[t]
  \centering
    \includegraphics[width=\linewidth]{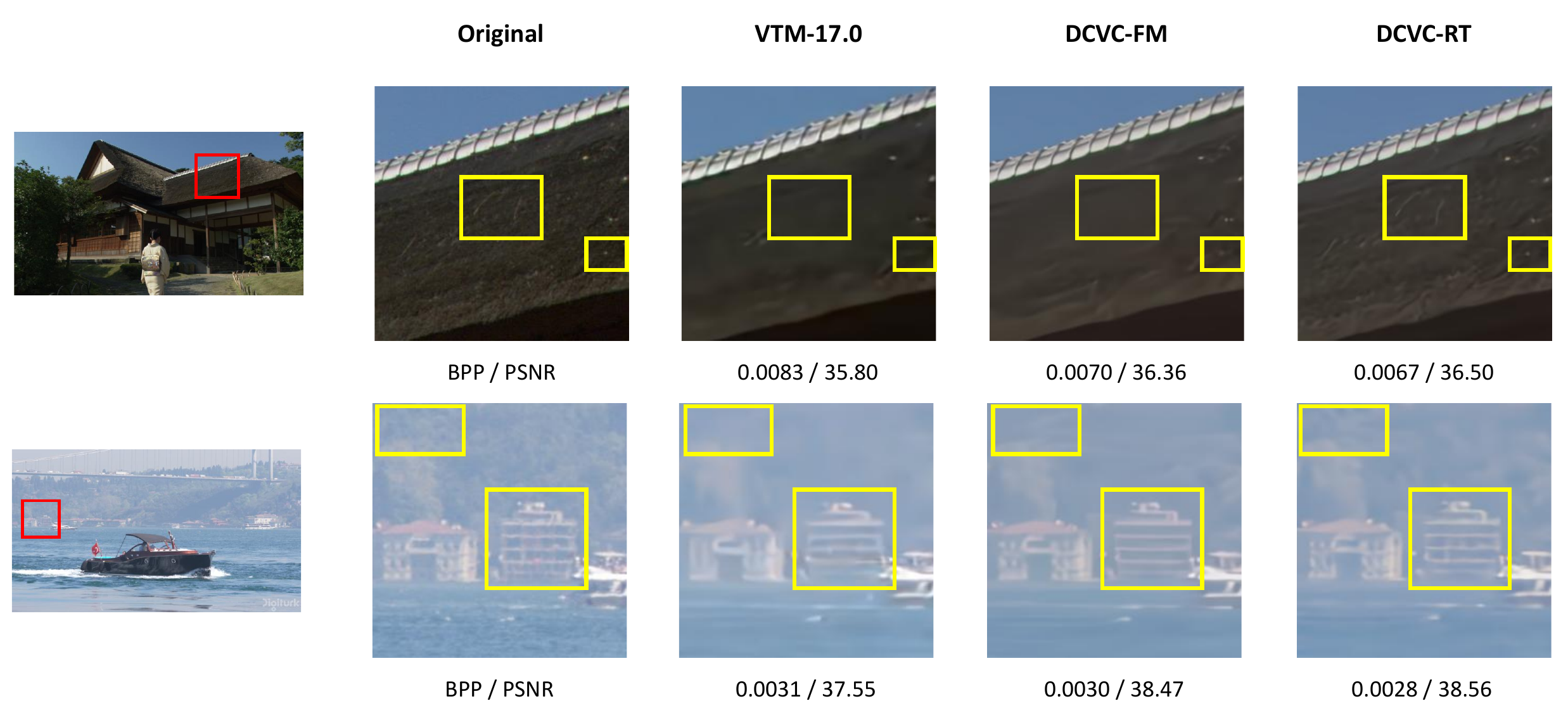}
    \vspace{-5mm}
    \caption{Visual comparison with VTM \cite{VTM} and DCVC-FM \cite{DCVC-FM}. \textit{Best viewed when zoomed in.}} 
  \label{fig:visual}
  \vspace{-3mm}
\end{figure*}

\begin{figure}[t]
  \centering
    \includegraphics[width=0.99\linewidth]{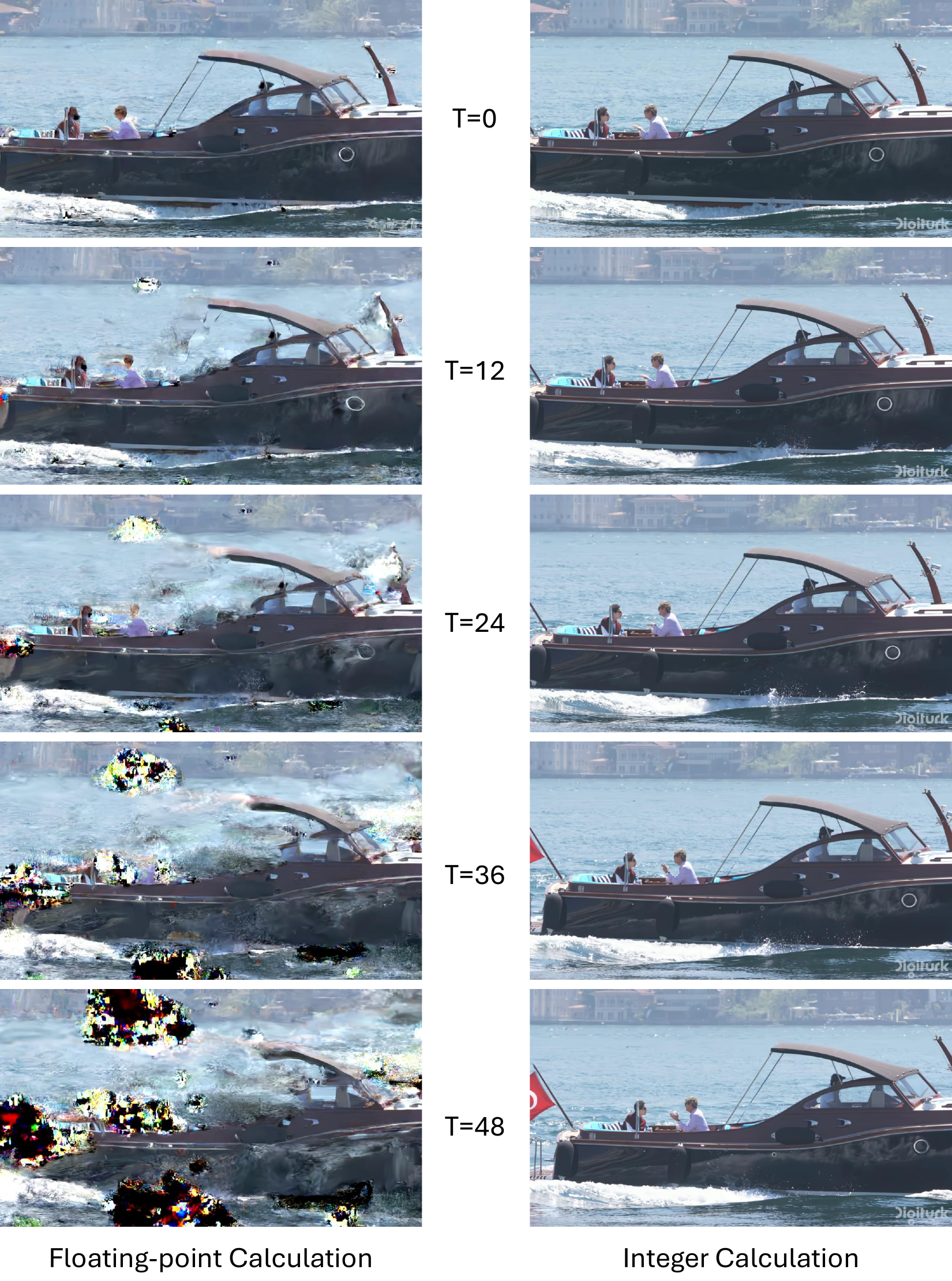}
    \vspace{-2mm}
    \caption{An example for cross-platform coding on \textit{YachtRide} sequence.  We perform encoding on an NVIDIA A100 GPU and perform decoding on an RTX 2080Ti. } 
  \label{fig:More_Cross_Device_Visual}
\end{figure}

\begin{table*}[t]
    \caption{BD-Rate (\%) comparison in YUV420 colorspace. All frames with intra-period=--1.}
    \vspace{-2mm}
    \centering
    \renewcommand\arraystretch{1.4}
    \setlength{\tabcolsep}{3.5pt}
    \resizebox{0.7\linewidth}{!}{
    \begin{tabular}{ l c c c c c c c }
        \toprule
        & UVG & MCL-JCV & HEVC B & HEVC C & HEVC D & HEVC E & Average \\
        \hline
        VTM-17.0    & $0.0$   & $0.0$   & $0.0$   & $0.0$   & $0.0$   & $0.0$   & $0.0$\\ 
        \hline
        DCVC-FM (fp16)     & $-16.8$ & $-8.0$ & $-15.4$ & $-30.2$ & $-37.5$ & $-20.2$ & $-21.3$ \\  
        \hline
        DCVC-RT (fp16)     & $-24.0$  & $-14.8$   & $-16.6$ & $-21.0$  & $-27.3$ & $-22.4$ & $-21.0$  \\
        \hline
        DCVC-RT (int16)     & $-21.0$  & $-12.3$   & $-14.8$ & $-20.0$  & $-26.4$ & $-15.0$ & $-18.3$ \\
        \hline
        DCVC-RT Large (fp16)     & $-31.1$  & $-22.1$   & $-27.7$ & $-32.1$  & $-37.7$ & $-34.0$ & $-30.8$  \\
        \hline
        DCVC-RT Large (int16)     & $-27.7$  & $-19.6$   & $-25.9$ & $-31.1$  & $-37.0$ & $-24.2$ & $-27.6$  \\
        \bottomrule
    \end{tabular}
	}
  \label{tab:sup-bd-int}
  \vspace{-1mm}
\end{table*}

\begin{table*}[t]
    \caption{BD-Rate (\%) comparison in RGB colorspace. All frames with intra-period=--1.}
    \vspace{-2mm}
    \centering
    \renewcommand\arraystretch{1.4}
    \setlength{\tabcolsep}{3.5pt}
    \resizebox{0.7\linewidth}{!}{
    \begin{tabular}{ l c c c c c c c }
        \toprule
        & UVG & MCL-JCV & HEVC B & HEVC C & HEVC D & HEVC E & Average \\
        \hline
        VTM-17.0    & $0.0$   & $0.0$   & $0.0$   & $0.0$   & $0.0$   & $0.0$   & $0.0$\\ 
        \hline
        HM-16.25    & $43.2$   & $49.5$   & $49.9$   & $45.2$   & $39.9$   & $47.7$   & $45.9$\\ 
        \hline
        DCVC-DC    & $9.2$   & $0.0$   & $14.9$   & $5.3$   & $-7.8$   & $87.7$   & $18.2$\\ 
        \hline
        DCVC-FM    & $-11.0$   & $-1.3$   & $-11.5$   & $-26.6$   & $-33.8$   & $-15.4$   & $-16.6$\\ 
        \hline
        DCVC-FM (fp16)    & $-10.4$   & $-1.1$   & $-11.2$   & $-26.5$   & $-33.7$   & $-12.1$   & $-15.8$\\ 
        \hline
        DCVC-RT (fp16)     & $-17.2$   & $-6.8$   & $-11.3$   & $-15.8$   & $-21.3$   & $-11.4$   & $-14.0$ \\
        \hline
        DCVC-RT (int16)     & $-13.1$   & $-3.9$   & $-9.2$   & $-14.8$   & $-20.4$   & $-3.2$   & $-10.8$ \\
        \hline
        DCVC-RT Large (fp16)     & $-25.0$  & $-14.8$   & $-22.9$ & $-27.5$  & $-32.4$ & $-24.4$ & $-24.5$  \\
        \hline
        DCVC-RT Large (int16)     & $-20.5$  & $-11.9$   & $-20.9$ & $-26.5$  & $-31.7$ & $-13.5$ & $-20.8$  \\
        \bottomrule
    \end{tabular}
	}
  \label{tab:sup-bd-rgb}
  \vspace{-1mm}
\end{table*}

\subsection{Interpolation for Arbitrary Rates}

In the proposed rate-adjustment module bank, we learn $64$ rate points to accommodate different rates within a single model. These modules include vectors for latent modulation and the factorized modules for coding $z$. In practice, we can expand the supported rate number to arbitrary larger by performing interpolation in the module bank. Specifically, to achieve a rate between the $i$-th and the $i+1$-th module, we perform linear interpolation between the two vectors to obtain an intermediate vector for latent modulation. For the factorized module, we directly select the nearest module for probability estimation. In Fig. \ref{fig:rate_adjustment_128}, we evaluate it to achieve $128$ different rates ($64$ original and $64$ interpolated) in a single model, and it  demonstrates a very smooth quality adjustment.

\subsection{Model Integerization Results}

Maintaining calculation consistency across different devices is critical for video codecs, as encoding-decoding inconsistencies can lead to errors in entropy coding and ultimately yield corrupted reconstructions. Fig. \ref{fig:More_Cross_Device_Visual} (left) shows how nondeterministic floating-point calculations can accumulate errors over time, producing visible artifacts after around 30 frames. In DCVC-RT, model integerization addresses this issue by facilitating model integerization and enforcing deterministic integer calculations. As shown in Fig. \ref{fig:More_Cross_Device_Visual} (right), this approach ensures cross-device consistency.

Tab. \ref{tab:sup-bd-int} and \ref{tab:sup-bd-rgb} present the rate-distortion performance of DCVC-RT in int16 mode. Compared to fp16, the int16 model exhibits only a minor BD-Rate decrease of approximately $3\%$, demonstrating its practicality and effectiveness for real-world applications.

\begin{figure*}[t]
  \centering
    \includegraphics[width=0.94\linewidth]{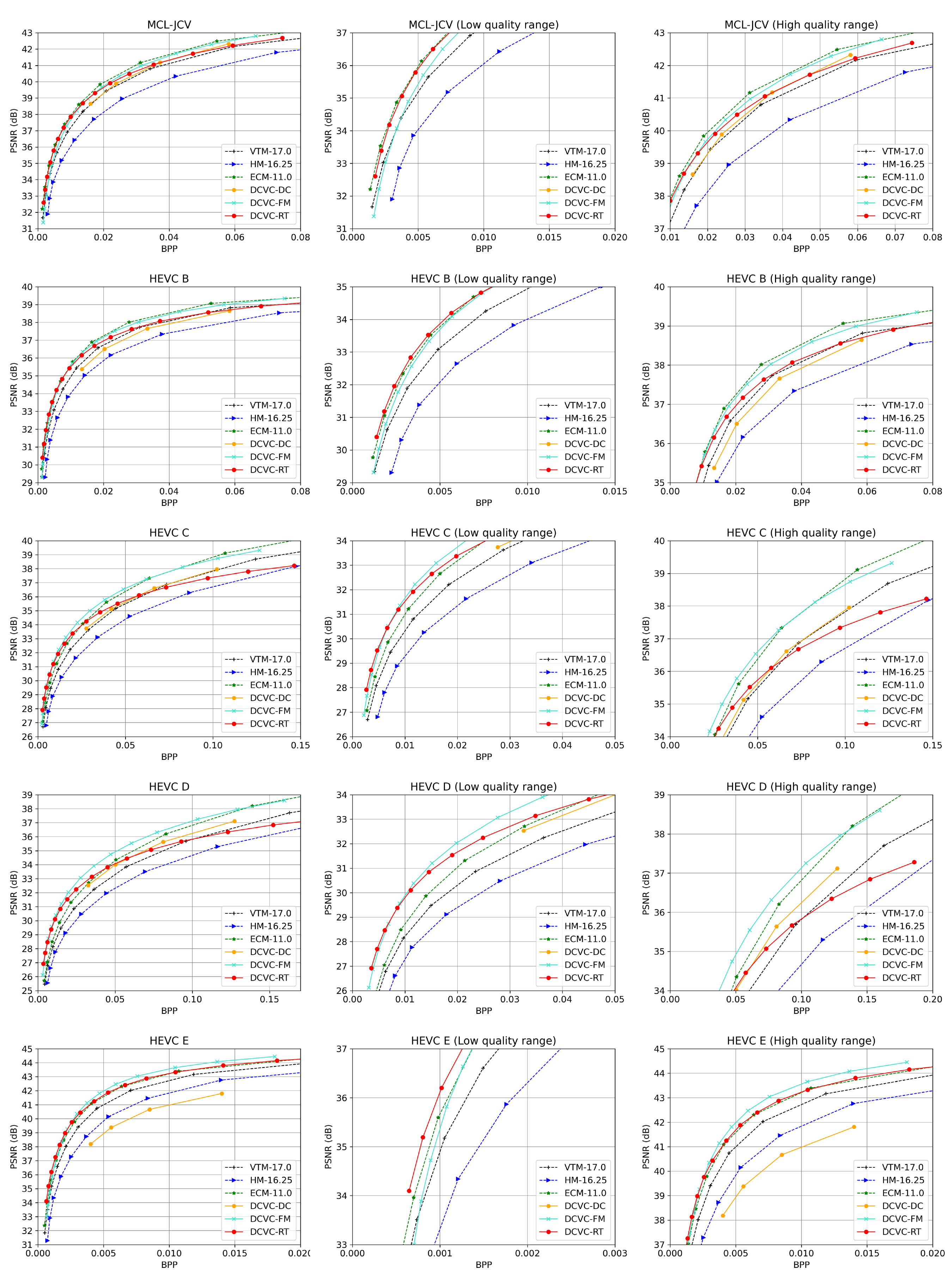}
    \caption{Rate-distortion curves for MCL-JCV and HEVC datasets. All frames are tested with intra-period=--1 in YUV420 colorspace. We show the whole quality range, relatively low quality range and relatively high quality range for each dataset. } 
  \label{fig:Full-RD-Curve}
\end{figure*}

\subsection{Results on RGB Colorspace}

In Tab. \ref{tab:sup-bd-rgb}, we present the BD-rate comparison for the RGB format under all frame intra period --1 settings. As depicted in the table, DCVC-RT achieves an average 14.0\% bits saving compared to VTM, which is comparable to 15.8\% of DCVC-FM. It demonstrate the high rate-distortion performance of DCVC-RT.

\subsection{Visual Comparison}

To further demonstrate the superiority of DCVC-RT, we also present the visual comparison with VTM \cite{VTM} and DCVC-FM \cite{DCVC-FM} in Fig. \ref{fig:visual}. From these examples, we can observe that DCVC-RT can reconstruct frames with more details and clearer structures, even at lower bitrates. We offer video visualizations and comparisons on our project page: \url{https://dcvccodec.github.io/} .

\begin{table}[t]
    \caption{Complexity analysis of DCVC-RT-Large. All methods are tested in fp16 mode on an A100 GPU. The coding speed are evaluated on 1080p videos. BD-Rate is calculated in YUV420 colorspace using VTM as anchor.
    }
    \vspace{-1mm}
    \centering
    \setlength{\tabcolsep}{6pt}
    \resizebox{0.9\linewidth}{!}{
    \begin{tabular}{ l | c | c | c  c}
        \toprule
        \multirow{2}{*}{Model} & Average & \multirow{2}{*}{MACs} & \multicolumn{2}{c}{Coding Speed}\\
        ~ & BD-Rate & ~ & Enc. & Dec. \\
        \midrule
        DCVC-FM & --21.3\% & 2642G & 5.0 fps & 5.9 fps \\
        \midrule
        DCVC-RT-Large & --30.8\% & 812G & 89.3 fps & 73.4 fps\\
        \midrule
        DCVC-RT & --21.0\% & 385G & 125.2 fps & 112.8 fps\\
        \bottomrule
    \end{tabular}
    }
  \label{tab:sup-complexity}
  \vspace{-3mm}
\end{table}
       
\subsection{Rate-Distortion Curves}
\label{sec:sup-rd}

In Fig. \ref{fig:Full-RD-Curve}, we present the rate-distortion curves across all tested datasets. On the low-quality range, DCVC-RT demonstrates comparable or superior performance to DCVC-FM and surpasses ECM, showcasing its exceptional performance. However, we observe a performance decline at high rates. This decrease can be attributed to the lightweight design of DCVC-RT model, which has limited capabilities compared to larger models. To examine this problem, we increase the model capacity of DCVC-RT in Section. \ref{sec:sup-scaling}.

\subsection{Performance for a Large Model}
\label{sec:sup-scaling}

In this paper, we primarily design a lightweight model to accelerate coding. However, DCVC-RT can be easily extended to a larger model (DCVC-RT Large) by increasing the number of channels and DCB blocks. As shown in Tables \ref{tab:sup-bd-int} and \ref{tab:sup-bd-rgb}, the enhanced model capacity significantly improves compression performance. On YUV420 colorspace, Our large model, DCVC-RT-Large, achieves an average BD-Rate of $-30.8\%$, outperforming $-21.3\%$ of the advanced large NVC model, DCVC-FM. On RGB colorspace, It achieves a BD-Rate of $-24.5\%$, compared to DCVC-FM's $-15.8\%$. Furthermore, the performance drop at high bitrates observed in the small model is effectively addressed in the large model, confirming the conclusion in Section. \ref{sec:sup-rd}. For high-range rate points (calculated from $qp=42$ to $qp=63$), using DCVC-FM as the anchor, DCVC-RT shows a BD-Rate loss of $9.3\%$ on UVG, whereas DCVC-RT-Large achieves a slightly better BD-Rate of $-1.3\%$. Despite its improved performance, DCVC-RT-Large maintains significantly lower complexity than DCVC-FM, as shown in Table. \ref{tab:sup-complexity}. Thanks to its efficiency-driven design, it achieves an encoding speed of approximately 90 fps. These results underscore the superiority of DCVC-RT in the rate-distortion-complexity trade-off.

\end{document}